\documentclass[a4paper,fleqn,usenatbib,onecolumn]{mnras}

\usepackage{newtxtext,newtxmath}

\usepackage[T1]{fontenc}
\usepackage{ae,aecompl}

\usepackage{graphicx}	
\usepackage{amsmath}	
\usepackage{amssymb}	

\title[Where are the solar magnetic poles?]{Where are the solar magnetic poles?}

\author[A. Pastor Yabar et al.]{
A. Pastor Yabar,$^{1,2}$\thanks{E-mail: apastor@iac.es (IAC)}
M. J. Mart\'{\i}nez Gonz\'alez,$^{1,2}$
M. Collados$^{1,2}$
\\
$^{1}$Instituto de Astrof\'{i}sica de Canarias, Calle V\'{\i}a L\'actea, s/n, E-38205 La Laguna, Tenerife, Spain\\
$^{2}$Universidad de La Laguna, E-38206 La Laguna, Tenerife, Spain\\
}

\date{Accepted XXX. Received YYY; in original form ZZZ}

\pubyear{2015}

\begin{document}
\label{firstpage}
\pagerange{\pageref{firstpage}--\pageref{lastpage}}
\maketitle

\begin{abstract}
   {Regardless of the physical origin of stellar magnetic fields -- fossil or dynamo induced -- an inclination angle between the magnetic and rotation axes is very often observed. Absence of observational evidence in this direction in the solar case has led to generally assume that its global magnetic field and rotation axes are well aligned. We present the detection of a monthly periodic signal of the photospheric solar magnetic field at all latitudes, and especially near the poles, revealing that the main axis of the Sun's magnetic field is not aligned with the surface rotation axis. This result reinforces the view of our Sun as a common intermediate-mass star. Furthermore this detection challenges and imposes a strong observational constraint to modern solar dynamo theories.}
\end{abstract}




\section{Introduction}
The study of the solar global magnetic field is important for two main reasons. It plays a critical role in the Sun--Earth interaction given that it directly affects the properties of the interplanetary magnetic field in which the Earth is embedded. The Sun is also the key object to understand the generation and evolution of magnetic fields in other late-type stars with different masses, ages and rotation speeds.\\
First studies of the solar global magnetic field in the mid-1990s led to the dynamo theory \citep{b1961, l1969}. Since then, this theory has evolved and is capable of explaining several global magnetic features of the Sun such as the 11-year sunspot cycle, the polarity inversion of the magnetic poles, the tilted emergence of sunspot groups, and the decreasing latitude of sunspot emergence during the 11-year activity cycle \citep{h1925}. From this theory, whose magnetic engine is solar rotation, it is assumed that the magnetic and the rotational axes of the Sun are aligned. Non-linear dynamo models are able to excite non-axisymmetric components of global magnetic field \citep{rss1988, m1999}. However these solutions still have some limitations and do not match both dynamic and magnetic features at the same time.\\
Previous studies of the solar global magnetic field  have found a high degree of alignment between rotational and magnetic axes for almost the full solar cycle \citep{w1993, drbh2012}. They claimed that both axes are misaligned only during the polarity reversal of the poles, where the dipole changes its orientation in about two years, and in sporadic excursions. For these studies, data from ground-based telescopes and from the \textit{Solar and Heliospheric Observatory} \citep[SOHO;][]{dfp1995} satellite have been used. These data lack of enough spatial resolution and magnetic sensitivity near the poles, where the global magnetic field is not affected by active regions. Accurate, long-term, synoptic magnetic field measurements, including the solar poles, are required in order to study properly the axisymmetryc component of the global magnetic field.

\section{Methods and Observations}

\begin{figure*}
\includegraphics[width=\textwidth]{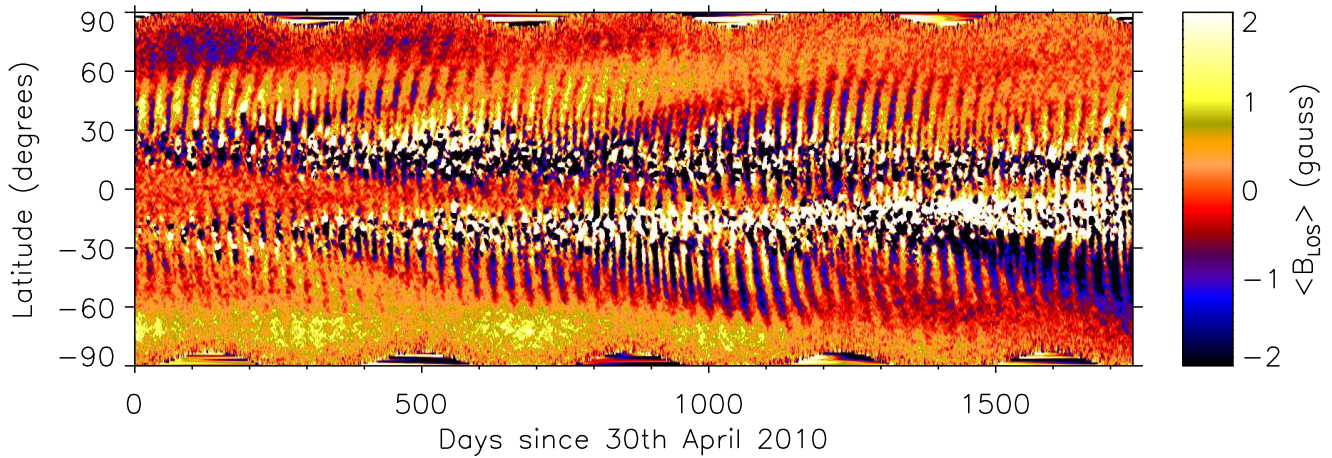}
 \caption{Time evolution of the line-of-sight magnetic field averaged over all solar longitudes at each latitude. Daily \textit{SDO}/HMI full-disc magnetograms have been used to sample a period of time from 2010 April 30 to 2015 January 24. Time is shown in the horizontal axis and the heliographic latitude in the vertical axis. The colour scale represents the average line-of-sight component of the magnetic field at each solar latitude. The latitude is discretized in a bin size of $1^{\circ}$.}
 \label{Fig. 1}
\end{figure*}
\begin{figure}
\begin{center}
\includegraphics[width=0.5\columnwidth]{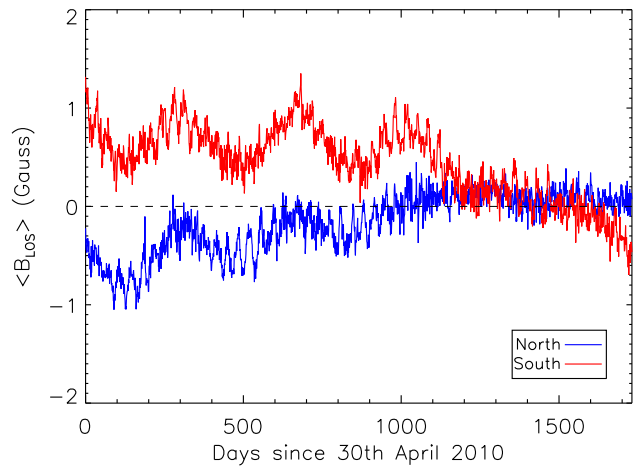}
 \caption{Time evolution of the line-of-sight magnetic field at the rotational poles. Each value of the longitudinal magnetic field has been obtained by averaging over all visible longitudes and between latitudes $70^{\circ}$ and $80^{\circ}$ for the south (red) and north (blue) poles.}
 \label{Fig. 2}
\end{center}
\end{figure}

Magnetograms --the projection of the magnetic field on to the observer's line of sight-- from the Helioseismic and Magnetic Imager \citep[HMI;][]{setal2012, s2etal2012} onboard the \textit{Solar Dynamics Observatory} \citep[SDO;][]{ptc2012} have been used in this Letter. One full-disc magnetogram per day has been used to sample the entire \textit{SDO} mission, around five years since 2010, covering the last minimum of activity and the present maximum. For each daily magnetogram the signal is averaged over all longitudes within $1^{\circ}$ latitudinal bins covering the whole visible solar disc. Fig. \ref{Fig. 1} shows the time evolution of the average longitudinal magnetic field for each latitude bin. Some well-known features of the global magnetic field along the activity cycle are easily discernible: The magnetogram signal at high latitudes (over 70$^{\circ}$) shows a decrease from the 2010 activity minimum to the 2015 activity maximum \citep{shlz2015}, which is expected to finally turn into a complete polarity reversal in the coming years. The equatorward drift of the emergence of active regions as the cycle evolves is also observable. Another prominent characteristic is the poleward migration of flux from the activity belt to the poles, which is explained by means of the meridional flow \citep{tmlh1982}.\\
Unexpectedly, a pervasive monthly oscillation is observed at all latitudes and for the whole time sequence. No systematic behaviour of the \textit{SDO} satellite reported reflects this periodicity \citep{hetal2014}. Interestingly, the signal is present even at high latitudes, near the solar poles, as shown in Fig. \ref{Fig. 2}, where the time sequence for the average line-of-sight magnetic field between $70^{\circ}$ and $80^{\circ}$ for both poles is plotted. This oscillation is only possible if a non-axisymmetric distribution of magnetic fields exists around the solar rotation axis. The inclination of the solar rotation axis with respect to the ecliptic makes this pattern more apparent at certain moments of the year, since each pole is best visible every six months while the opposite pole is hidden. Hence, every year the signal is more clearly detected when those latitudes are optimally observed. The amplitude of this fluctuation is the same during the activity minimum and maximum, when the polar magnetic field reaches its maximum and minimum values, respectively. The yearly variation of the inclination of the solar rotation axis is the reason for the large amplitude periodic fluctuation observed in Fig. \ref{Fig. 2}. The decreasing/increasing trend of the high-latitude southern/northern magnetic field strength is also apparent, situating the polar reversal around day 1400/1000 from the beginning of the time series.\\
\begin{figure*}
\includegraphics[width=\textwidth]{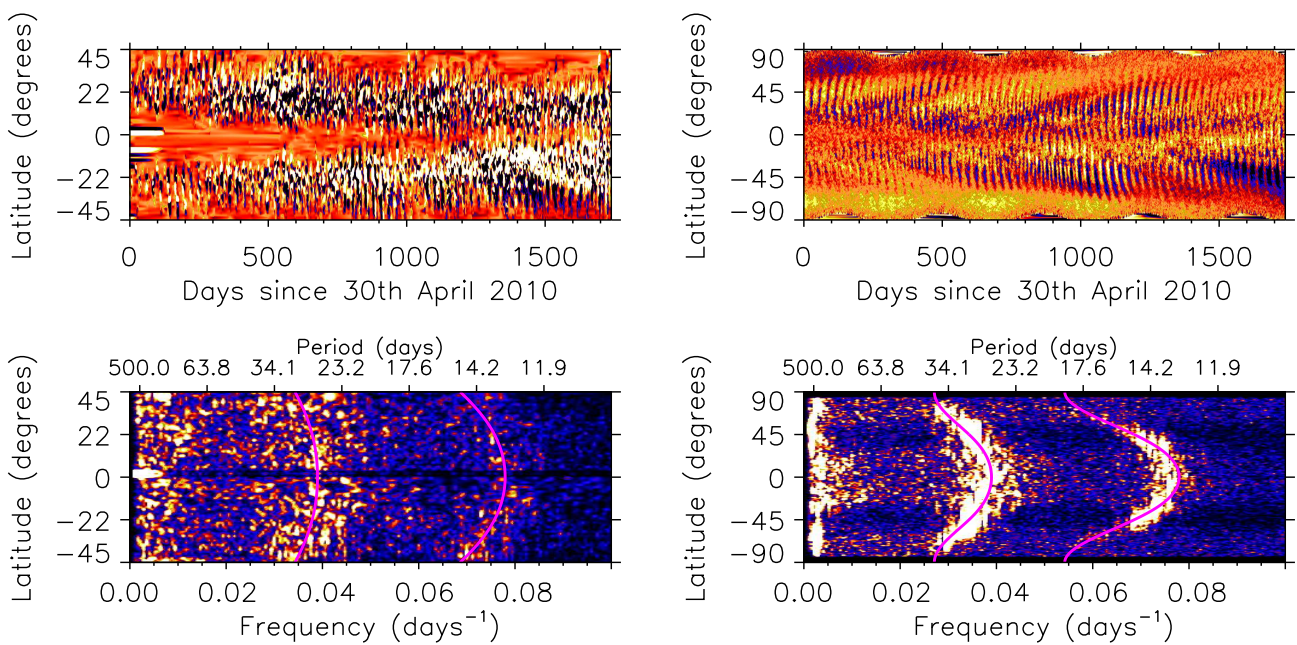}
 \caption{Contributions to the average line-of-sight magnetic field of active and quiet Sun. Same magnitude as in Fig. \ref{Fig. 1} separated in two components: active regions (left-hand panel) and quiet Sun (right-hand panel). The colour scale is as in Fig. \ref{Fig. 1}. As expected, active regions are only found between $\pm$45$^{\circ}$. Bottom panels display the power spectra in units of G$^2$ day normalized to the total power distributed over all frequencies. For the quiet Sun the power spectrum between latitudes 83 and 90 degrees (north and South) has not been calculated since the time series is not continuous due to the tilt angle between the ecliptic and the solar rotation axis. The overplotted purple lines, represent the standard synodic differential rotation profile, following the expression: $\omega=A+B\sin^2\lambda+C\sin^4\lambda$, where $\omega$ is the angular velocity and $\lambda$ the latitude. The constants have values \citep{su1990} $A=2.972$, $B=-0.484$ and $C=-0.361$.}
 \label{Fig. 3}
\end{figure*}

Non-axisymmetric configurations of the magnetic field, hence leading to monthly periodic signals, are likely in the activity belt, since opposite polarities due to active regions do not cancel out perfectly in latitude bins. To understand the role of active regions and quiet Sun, for each time step, a mask has been created to isolate the contribution of active and quiet components. In order to classify each point of the solar surface, a square of $2.5\times2.5$ arcsec$^2$ has been taken centred on it. If the mean absolute value of the line-of-sight magnetic field over this square is above/below 20 G the point is sorted as active/quiet region. Fig. \ref{Fig. 3} displays the time evolution of the average line-of-sight magnetic field for each component, and its associated power spectrum. It is remarkable that both components behave differently: Active regions are contained within $\pm$ $45^{\circ}$ in latitude and they drift to the equator as the cycle evolves. Most of the surface is quiet Sun most of the time and its power spectrum shows an eye-catching concentration of the power highly located at differential rotation frequencies (both for the fundamental mode and the first harmonic). The observational curve of the synodic differential rotation profile \citep{su1990} is overplotted in purple colour to point out the doubtless relation of the detected periodic signal with the solar surface rotation. Active regions show, however, a much more disperse power spectrum. This is expected since they emerge randomly in longitude and are not long-lived enough to have temporal coherence to concentrate the power in the rotation frequencies.\\

\section{Conclusions}

It is possible to conjecture about potential scenarios that are compatible with this undoubted rotational modulation of the magnetic signal at all latitudes. It is generally accepted that the average magnetic field at the solar poles evolves in time as a consequence of the residual magnetic field of active regions that is advected by the meridional flow from the activity belt up to higher latitudes \citep{wsn1991, dc1999}. It is to be noted that the amount of positive flux which is advected to the north pole is larger than the negative flux advected to the south pole at the beginning of the temporal series, as seen in Figs. \ref{Fig. 1} and \ref{Fig. 3}, leading to a late reversal of the latter. The detected rotation-induced magnetic oscillation can only be explained if this advection occurs at preferential longitudes and at all latitudes, and flux is advected such that half of the regions around the poles have a slightly larger magnetic field than the opposite half. This coherent advection at all latitudes seems unlikely to happen during the five years covered by the data. In addition, as already pointed out, the amplitude of the rotation modulation seems to be independent of the cycle phase (see Fig. \ref{Fig. 2}) and consequently independent of the active regions present at each moment on the solar surface. This result indicates that the observed periodicity is an intrinsic property of the solar global magnetic field, not directly related to the activity cycle itself.\\
This conclusion leads to a scenario based on a dynamo model in which the global field is not axisymmetric with respect to the rotation axis \citep{gcs2010, pg1994, rgwp1996}. A simple explanation is obtained with a bipolar field that is tilted a certain angle to the surface rotation axis, leading to a natural oscillation of the observed magnetic signal with solar rotation. The fact that the amplitude of the monthly oscillation remains roughly constant during the whole time series --which spans a minimum and a maximum of activity-- suggests that the tilt does not vary much. A possible explanation for the observed misalignment is a variation of the rotation axis with depth in the convection zone. This variation would break the symmetry and lead to an apparent shift of the magnetic poles on the surface. Helioseismology may help to obtain observational evidences of this varying orientation with depth of the rotation axis. The dependence of the rotation speed with depth is obtained after the splitting of the oscillation modes in a number of distinct frequencies. This is feasible since different modes propagate preferentially in different layers, some of them closer to the surface and some of them getting deeper into the sun's interior. The information on the rotation axis is more difficult to achieve, though. Not only the mode splitting is required, but also the relative amplitudes of the split components need to be accurately measured. The determination of these relative amplitudes represent a hard task for helioseismoly and it remains to be determined whether present space data have the necessary sensitivity for this measurement or if, on the contrary, more precise space instrumentation is required.\\
The misalignment between surface rotation and the global magnetic field of the Sun has a number of additional important consequences. The Sun is no longer a peculiar star and the detected misalignment has to be taken as an observational constraint for the theories of generation of stellar magnetic fields. Current widely accepted solar dynamo theories do not consider the possibility to get such a tilt between the rotation and the magnetic field \citep{c2010, dc1999, drbh2012}. This is a result directly derived of the assumption of a common rotation axis with depth. The main result presented in this Letter suggests that this assumption may need to be relaxed. It would not be surprising that a varying rotation axis with depth leads to instabilities that generate in a natural way a magnetic cycle.  The misalignment must also have an impact on the solar wind \citep{pg1994, rgwp1996}, the interplanetary magnetic field and the Earth's magnetosphere and the impact the Sun has on them on a monthly temporal scale. In fact, strong inclinations of the heliospheric current sheet (HCS) have already been reported in the past \citep{nrp2008, h1977, w1993, dxf2008} which may be related to the existence of an inclined solar magnetic dipole. Quasiperiodic 27-day variations of the interplanetary magnetic field ($\pm$5nT) and solar wind speed (400 $km\ s^{-1}$) have been detected \citep{ff2012, nsrfv2000} and can be interpreted as the result of a tilt angle of the HCS with respect to the solar rotation axis. Also, short-term fluctuations of cosmic rays flux with a 27-d period have been reported \citep{ma2013}. These fluctuations may be directly related to the modulation of the heliospheric magnetic field as a result of the misalignment reported here, which opens a new window for short-term climate studies in a scenario in which the solar magnetic field, cosmic rays and climate variations are the main actors.\\
From an observational point of view, a larger effort is needed to have longer time series with better polarimetric sensitivity to determine the weak polar signals. We have tried to extend our analysis to a larger temporal window by applying the same study to \textit{SOHO}/MDI \citep{s3etal1995} data. However, the weakness of the signals prevents us from using them to get a longer temporal series. Further analyses would be possible if the \textit{SDO} mission continues recording data during the next years at least until the new minimum of activity --near 2021--, and cover a complete 11-year cycle. Extremely valuable data are also expected from the \textit{Solar Orbiter Mission} \citep{mmsg2013} since this satellite is going to observe the rotational solar poles from outside the ecliptic and simultaneously detect solar wind particles close to their origin place.\\

\section*{Acknowledgements}

The uthors are especially grateful to Rafael Manso Sainz and to Allan Sacha Brun for very interesting discussions. This work is based on data courtesy from NASA/\textit{SDO} and the HMI science team. Financial support by the Spanish Ministry of Economy and Competitiveness and the European FEDER Fund through project AYA2010-18029 (Solar Magnetism and Astrophysical Spectropolarimetry) is gratefully acknowledged.




\bibliographystyle{mnras}
\bibliography{biblio} 








\bsp	
\label{lastpage}
\end{document}